\documentclass[twocolumn,prb,aps,amsfonts,showpacs]{revtex4}

\usepackage{graphicx}

\begin{document}
\draft
\preprint{}

\newcommand{\1}{{\bf \scriptstyle 1}\!\!{1}}
\newcommand{\I}{{\rm i}}
\newcommand{\p}{\partial}
\newcommand{\D}{^{\dagger}}
\newcommand{\bs}{{\bf s}}
\newcommand{\bx}{{\bf x}}
\newcommand{\br}{{\bf r}}
\newcommand{\bk}{{\bf k}}
\newcommand{\bv}{{\bf v}}
\newcommand{\bp}{{\bf p}}
\newcommand{\bu}{{\bf u}}
\newcommand{\bA}{{\bf A}}
\newcommand{\bB}{{\bf B}}
\newcommand{\bE}{{\bf E}}
\newcommand{\bF}{{\bf F}}
\newcommand{\bG}{{\bf G}}
\newcommand{\bI}{{\bf I}}
\newcommand{\bK}{{\bf K}}
\newcommand{\bL}{{\bf L}}
\newcommand{\bP}{{\bf P}}
\newcommand{\bQ}{{\bf Q}}
\newcommand{\bS}{{\bf S}}
\newcommand{\bH}{{\bf H}}
\newcommand{\balpha}{\mbox{\boldmath $\alpha$}}
\newcommand{\bsigma}{\mbox{\boldmath $\sigma$}}
\newcommand{\btau}{\mbox{\boldmath $\tau$}}
\newcommand{\bSigma}{\mbox{\boldmath $\Sigma$}}
\newcommand{\bOmega}{\mbox{\boldmath $\Omega$}}
\newcommand{\bpi}{\mbox{\boldmath $\pi$}}
\newcommand{\bphi}{\mbox{\boldmath $\phi$}}
\newcommand{\bnabla}{\mbox{\boldmath $\nabla$}}
\newcommand{\bmu}{\mbox{\boldmath $\mu$}}
\newcommand{\bepsilon}{\mbox{\boldmath $\epsilon$}}

\newcommand{\iLambda}{{\it \Lambda}}
\newcommand{\cA}{{\cal A}}
\newcommand{\cD}{{\cal D}}
\newcommand{\cF}{{\cal F}}
\newcommand{\cL}{{\cal L}}
\newcommand{\cH}{{\cal H}}
\newcommand{\cI}{{\cal I}}
\newcommand{\cM}{{\cal M}}
\newcommand{\cO}{{\cal O}}
\newcommand{\cR}{{\cal R}}
\newcommand{\cU}{{\cal U}}
\newcommand{\cT}{{\cal T}}

\newcommand{\be}{\begin{equation}}
\newcommand{\ee}{\end{equation}}
\newcommand{\bea}{\begin{eqnarray}}
\newcommand{\eea}{\end{eqnarray}}
\newcommand{\beqa}{\begin{eqnarray*}}
\newcommand{\eeqa}{\end{eqnarray*}}
\newcommand{\nn}{\nonumber}
\newcommand{\DD}{\displaystyle}

\newcommand{\ba}{\left[\begin{array}{c}}
\newcommand{\baa}{\left[\begin{array}{cc}}
\newcommand{\baaa}{\left[\begin{array}{ccc}}
\newcommand{\baaaa}{\left[\begin{array}{cccc}}
\newcommand{\ea}{\end{array}\right]}

\title{Theory of Umklapp-assisted recombination of bound excitons in Si:P}

\author{Michael N.~Leuenberger and L.~J.~Sham
}
\affiliation{Department of Physics, University of California San Diego, La Jolla, CA 92093}

\date{\today}

\begin{abstract}
We present the calculations for the oscillator strength of the recombination of excitons bound to phosphorous donors in silicon. We show that the direct recombination of the bound exciton cannot account for the experimentally measured oscillator strength of the no-phonon line. Instead, the recombination process is assisted by an umklapp process of the donor electron state.
We make use of the empirical pseudopotential method to evaluate the Umklapp-assisted recombination matrix element in second-order perturbation theory. Our result is in excellent agreement with the experiment. We also present two methods to improve the optical resolution of the optical detection of the spin state of a single nucleus in silicon.
\end{abstract}


\maketitle

\section{Introduction}

Proposals on quantum computing in semiconductors have recently attracted a great deal of attention.\cite{Ziese,Awschalom} The main idea is to use the electron spin of quantum dots in semiconductors\cite{Loss} or the nuclear spin of shallow donors in silicon\cite{Kane} as a qubit for quantum information processing. A complete quantum computation consists of, besides the single- and two-qubit operations, the initialization and the readout of the qubits. While the initialization and the readout of the electron spin in quantum dots have been successfully demonstrated experimentally,\cite{Kroutvar,Dutt} it remains an experimental challenge to read out the nuclear spin of a donor electron in silicon.

It has recently been proposed that the photoluminescence of excitons bound to phosphorous donors can be used to detect the spin state of a single donor nucleus in silicon.\cite{Fu} This scheme for optical readout could render the recent quantum computing proposal using conditional NMR and ESR pulses feasible.\cite{BermanPRL86,BermanPRL87} Although experiments have shown that the recombination of the bound exciton follows strong optical selection rules,\cite{Sauer} it has been unclear what physical process is responsible for the no-phonon line of the optical recombination of the bound exciton. In contrast to the phonon-assisted recombination, the no-phonon line represents the recombination process without phonon assistance.
A shell model that accounts for the selection rules has been proposed in Ref.~\onlinecite{Kirczenow}.
The shell model has been later improved in Ref.~\onlinecite{Chang1980,Chang1982} by a Hartree-Fock calculation that takes the multivalley character of indirect bandgap semiconductors into account and therefore is in good agreement with the measured fine-structure excitation spectrum of the bound exciton complex.\cite{Lightowlers}
The Hartree-Fock calculation fails to predict binding energies. This problem was overcome by a density-functional calculation that takes the correlation energy into account.\cite{Pfeiffer}
However, neither model gives a satisfactory physical description of the recombination process for the no-phonon line and thus is unable to quantitatively reproduce the measured oscillator strength $f_{\rm exp}$ in Ref.~\onlinecite{Dean}.
Our calculations show that the probability for direct recombination of the bound exciton is negligibly small. 
Here we present a physical model of the recombination process that accurately reproduces the oscillator strength $f_{\rm exp}$ of the no-phonon line. In our model the recombination of the exciton is assisted by the Umklapp-process of the donor electron. In Sec.~\ref{model} we give a detailed description of our model, where we introduce the Coulomb scattering of the bound and the donor electron.
For the evaluation of the recombination matrix element, we make use of the empirical pseudopotential method\cite{Cohen,Chelikowsky1974,Chelikowsky1976} to calculate the bandstructure of silicon with 137 reciprocal lattice vectors at 100 points inside the first Brillouin zone in X direction. This technique is applied to our model in Sec.~\ref{bandstructure}.
Then we use the resulting Bloch states to calculate the oscillator strength of the Umklapp-assisted recombination in second-order perturbation theory, which is shown in Sec.~\ref{fosc}.
In Sec.~\ref{readout} we present two methods based on optically detected magnetic resonance (ODMR) to improve the resolution of the readout of a nuclear spin of a donor electron in silicon.

\section{Physical model}
\label{model}

Our physical model of the Umklapp-assisted recombination is shown in Fig.~\ref{Wavefunctions_umklapp}. First the bound electron with wavevector $\bk_{\rm B}$ is scattered via the Coulomb potential 
\be
V_{\rm C}=\frac{q^2e^{-r\sqrt{\xi_l^2+2\xi_t^2}}}{4\pi\epsilon r}
\ee
off the donor electron with wavevector $\bk_{\rm D}$, thereby conserving the momenta, i.e. 
$\bk_{\rm B}+\bk_{\rm D}=\bk_{\rm B}'+\bk_{\rm D}'$. $\epsilon_0=8.854\times 10^{-12}$ F/m is the dielectric constant of the vacuum, and $\epsilon_r=11.56$ is the relative dielectric constant of silicon, which are combined to $\epsilon=\epsilon_0\epsilon_r$.\cite{Bolivar}
So the refractive index of silicon is about $\nu=\sqrt{\epsilon_r}=3.4$. $\xi_l=4m_lk_F q^2/\pi\hbar^2=1.9\times 10^{9}$ m$^{-1}$ and $\xi_t=4m_tk_F q^2/\pi\hbar^2=4.0\times 10^{8}$ m$^{-1}$ are the Thomas-Fermi screening lengths in longitudinal and transverse direction, where $k_F=(3\pi^2n_0)^{1/3}$ is the Fermi wavevector and $n_0=1/L^3$ is the density of the donor electrons.\cite{Ashcroft} The distance between the donor electrons is about $L=4$ nm.\cite{Hawley} The longitudinal mass in silicon is $m_l=0.9163m_e$, where $m_e=9.1095\times 10^{-31}$~kg is the bare electron mass. The transverse mass in silicon is $m_t=0.1905m_e$.
After the Umklapp process the bound electron with wavevector $\bk_{\rm B}'$ recombines with the bound hole with wavevector $\bk_{\rm h}'=\bk_{\rm B}'$ via the electron-photon interaction
\be
V_{\rm r}=\frac{q}{m_t}A_\bot p_\bot,
\ee
which results in the emission of a photon that can be seen in a photoluminescence experiment.
So the total Hamiltonian describing our system is given by
\be
\cH=\sum_i\frac{p_i^2}{2m_i^*}+V_{\rm C}+V_{\rm r},
\label{Hamiltonian}
\ee
where the first term is the kinetic energy of the bound electron, the donor electron, and the bound hole with the effective masses $m_i^*$, which takes the lattice potential via the empirical pseudopotential method into account (see App.~\ref{pseudo}).

\begin{figure}[htb]
\includegraphics[width=8.5cm]{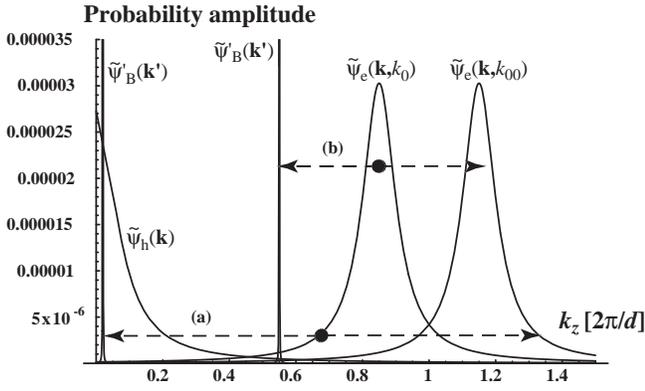}
\caption[]{Although the Coulomb scattering amplitude $M_C$ has a maximum at $k_u=0.55k_{\rm si}$ (see Fig.~\ref{MC}), which arises from the scattering process (b), the largest contribution to the Umklapp-assisted recombination comes from the scattering process (a), because the overlap of the unbound scattered electron and the hole is largest around $k_u=0$ (see Fig.~\ref{Mr}).}
\label{Wavefunctions_umklapp}
\end{figure}

The transition amplitude of the Umklapp-assisted recombination is given by
\be
M_{\rm tot}=\frac{1}{k_0}\int_0^{k_0}dk_{\rm u}M(k_{\rm u})=\frac{1}{k_0}\int_0^{k_0}dk_{\rm u}\frac{M_{\rm C}M_{\rm r}}{E_{\rm donor}-E_{\rm c}(k_{\rm u})}
\label{matrixelement}
\ee 
in second-order perturbation theory. $E_{\rm donor}$ is the energy level of the donor electron and $E_{\rm c}(k_{{\rm B}0}')$ is the energy of the conduction band. In order to calculate the Umklapp-assisted scattering matrix element $M_{\rm C}$ and the recombination matrix element $M_{\rm r}$, we need to identify the initial, intermediate, and final states (see Fig.~\ref{Wavefunctions_umklapp} and App.~\ref{matrix}).
Since the donor and the bound electron are in a spin singlet state $(\left|\uparrow_{\rm B}\downarrow_{\rm D}\right>-\left|\downarrow_{\rm B}\uparrow_{\rm D}\right>)/\sqrt 2$, the orbital wavefunction is symmetric in the valley combinations 
$j=\pm\bk_{x0},\pm\bk_{y0},\pm\bk_{z0}$, i.e.,
\be
\psi_{\rm BD}=\sum_{j}F_{{\rm B}j}\phi_{{\rm B}j}F_{{\rm D}j}\phi_{{\rm D}j},
\ee
where $\pm\bk_{x0}=(\pm k_0,0,0)$, $\pm\bk_{y0}=(0,\pm k_0,0)$, $\pm\bk_{z0}=(0,0,\pm k_0)$, and e.g.,
\be
F_{{\rm e}\bk_{z0}}(\br)=F_{{\rm B}\bk_{z0}}(\br)=F_{{\rm D}\bk_{z0}}(\br)=\frac{1}{\sqrt{a^2b}}e^{-\frac{|x|+|y|}{a}}e^{-\frac{|z|}{b}}
\ee
is the envelope function of the bound and the donor electron with $a=25.1$ {\AA} and $b=14.4$ {\AA} (see Refs.~\onlinecite{Kohn,Koiller}), and
\be
\phi_{{\rm e}\bk_{z0}}(\br)=\phi_{{\rm B}\bk_{z0}}(\br)=\phi_{{\rm D}\bk_{z0}}(\br)=u_{{\rm c}z}(\br)e^{ik_0z}
\ee
is the Bloch wavefunction.
$k_0=0.85k_{\rm si}$ is the distance from the $\Gamma$ point to the minimum of the conduction band in X direction (see next section).
$k_{\rm si}=2\pi/d$ is the reciprocal lattice vector of silicon with a lattice constant of $d=5.43$ \AA.
Similarly the envelope function of the hole state is given by
\be
F_{{\rm h}}(\br)=\frac{1}{\sqrt{c^3}}e^{-\frac{|x|+|y|+|z|}{c}}
\ee
with $c=4\pi\epsilon\hbar^2/m_{\rm hh}q^2$, where $m_{\rm hh}=0.523m_{\rm e}$. The Bloch wavefunction of the hole is
\be
\phi_{{\rm h}}(\br)=u_{{\rm v}}(\br).
\ee

\begin{figure}[htb]
\includegraphics[width=8.5cm]{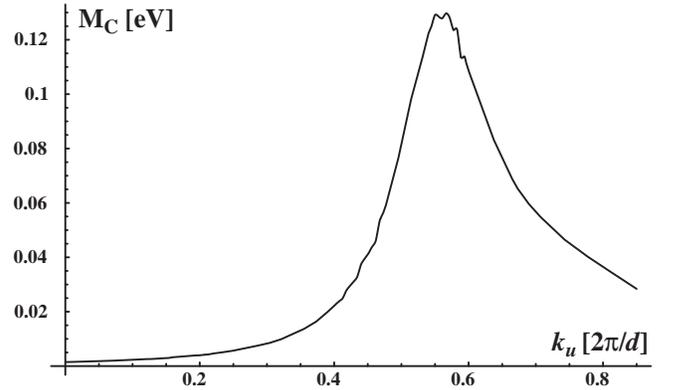}
\caption[]{The Coulomb scattering amplitude between the bound and the donor electron has a maximum at $k_u=0.55k_{\rm si}$.}
\label{MC}
\end{figure}

For the Coulomb scattering matrix element we obtain
\bea
M_{\rm C} & = & \tau_\bot^2\zeta \frac{q^2}{\epsilon}\int dk_{{\rm B}z}\int dk_{{\rm D}z}
\tilde{\psi}{'}_{{\rm D}0}^*(k_{{\rm B}z}-k_u+k_{{\rm D}z}) \nn\\
&  & \times 
\frac{1}{\left(k_{{\rm B}z}-k_u\right)^2+\xi^2} \tilde{\psi}_{{\rm B}0}(k_{{\rm B}z})\tilde{\psi}_{{\rm D}0}(k_{{\rm D}z}),
\eea
which is shown in Fig.~\ref{MC}. The maximum of $M_{\rm C}$ is located at $k_u=0.55k_{\rm si}$, which is due to the scattering process (b) shown in Fig.~\ref{Wavefunctions_umklapp}.
For the recombination matrix element we obtain
\be
M_{\rm r}=\frac{\hbar qA_\bot}{m}\int d^3k_{\rm h}'\tilde{\psi}_{\rm h}{'}^*(\bk_{\rm h}')k_{{\rm h}\bot}'
\tilde{\psi}_{\rm B}'(\bk_{\rm h}'),
\ee
which is shown in Fig.~\ref{Mr}. $M_{\rm r}$ has a maximum at $k_u=0$.
The analytical derivations of $M_{\rm C}$ and $M_{\rm r}$ can be found in App.~\ref{matrix}.
Next we need to calculate the bandstructure of silicon, which is presented in the next section.

\section{Bandstructure of silicon}
\label{bandstructure}

\begin{figure}[htb]
\includegraphics[width=8.5cm]{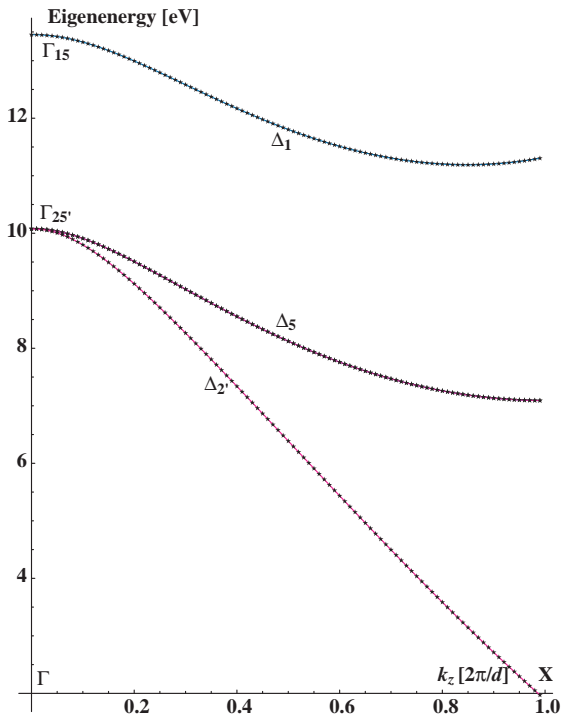}
\caption[]{Bandstructure of silicon in the direction of the X point. The minimum of the conduction band with $\Delta_1$ symmetry is located at $k_z=0.85k_{\rm si}$.}
\label{Bandstructure}
\end{figure}

We choose the empirical pseudopotential method to calculate 
the bandstructure of silicon.\cite{Cohen,Chelikowsky1974,Chelikowsky1976}
A brief review of this method is given in App.~\ref{pseudo}.
We need to solve the Schr\"odinger equation for the pseudowavefunction, i.e.,
\be
\frac{\hbar^2(\bk+\bG)^2}{2m_e}u_\bG+\sum_{\bG'}V_0(|\bG-\bG'|)u_{\bG'}=Eu_\bG.
\ee
We diagonalize it using the 137 $\bG$-vectors shown in Tab.~\ref{Gs} and the form factors $V_S(G=\sqrt{3})=-3.049$ eV, $V_S(G=\sqrt{8})=0.750$ eV, and $V_S(G=\sqrt{11})=0.985$ eV.

\begin{table}[htb]
\begin{tabular}{|c|r|c|r|r|}
\hline
$\bG$ & $\bG^2$ & direction & no. & sum \\\hline
(000) & 0 & $\Gamma$ & 1 & 1 \\
(111) & 3 & 2L & 8 & 9 \\
(200) & 4 & 2X & 6 & 15 \\
(220) & 8 & 2K & 12 & 27 \\
(311) & 11 & 2L+2X & 24 & 51 \\
(222) & 12 & 4L & 8 & 59 \\
(400) & 16 & 4X & 6 & 65 \\
(331) & 19 & 2L+2K & 24 & 89 \\
(420) & 20 & 2X+2K & 24 & 113 \\
(422) & 24 & 4L+2X & 24 & 137 \\\hline
\end{tabular}
\caption{$\bG$-vectors used in the empirical pseudopotential calculation.}
\label{Gs}
\end{table}

We calculate the bandstructure of silicon for 100 points in $\bk$-space in the direction of the X point. The bandstructure is shown in Fig.~\ref{Bandstructure}. For our Umklapp-assisted recombination the following symmetries are important: the conduction (valence) band at the $\Gamma$ point has the symmetry $\Gamma_{25'}$ ($\Gamma_{15}$), which transforms as $\{yz,xz,xy\}$ ($\{x,y,z\}$). The conduction band at the X point has the symmetry $\Delta_1$, which transforms as $\{z\}$. So the initial bound electron state before the recombination has the symmetry $\Delta_1$ and the hole has the symmetry $\Gamma_{15}$. Since the spin-orbit splitting at the $\Gamma$ point is $\Delta_{\rm so}=0.044$ eV (see Ref.~\onlinecite{Bassani}), the heavy-hole states including spin are represented by
\bea
\left|J=\frac{3}{2};M=\frac{3}{2}\right> & = & \frac{1}{\sqrt 2}\left(\left|yz\right>+i\left|xz\right>\right)\left|\uparrow\right>, \nn\\
\left|J=\frac{3}{2};M=\frac{1}{2}\right> & = & \sqrt{\frac{2}{3}}\left|xy\uparrow\right>-\frac{1}{\sqrt 6}\left|(y+ix)z\downarrow\right>, \nn\\
\left|J=\frac{3}{2};M=-\frac{1}{2}\right> & = & \sqrt{\frac{2}{3}}\left|xy\downarrow\right>+\frac{1}{\sqrt 6}\left|(y-ix)z\uparrow\right>, \nn\\
\left|J=\frac{3}{2};M=-\frac{3}{2}\right> & = & \frac{1}{\sqrt 2}\left(\left|yz\right>-i\left|xz\right>\right)\left|\downarrow\right>.
\eea
That is the origin of the selection rules found experimentally in Ref.~\onlinecite{Sauer}.

\begin{figure}[htb]
\includegraphics[width=8.5cm]{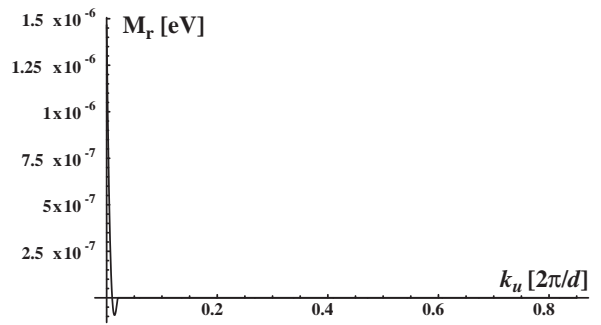}
\caption[]{The recombination amplitude for the unbound scattered electron is largest around $k_u=0$. The oscillator strength for the recombination of the scattered electron is $f_{\rm rec}(k_u=0)=0.023$.}
\label{Mr}
\end{figure}

\section{Oscillator strength}
\label{fosc}

In textbooks such as Ref.~\onlinecite{Cohen-Tannoudji1977} the oscillator strength is defined as
\be
f=\frac{2}{mE_{ni}}\left|\left<n\right|p\left|i\right>\right|^2,
\ee
where $E_{ni}$ is the energy difference between the initial $\left|i\right>$ and the final state $\left|n\right>$.
For our Umklapp-assisted recombination the oscillator strength is thus given by
\be
f_{\rm tot}=\frac{2}{m_{\rm opt}E_{\rm gap}}\left|\frac{g_{\rm hh}g_sM_{\rm tot}}{\frac{q}{m_t}A_\bot}\right|^2,
\ee
where $E_{\rm gap}=1.1$ eV is the indirect bandgap energy, $m_{\rm opt}=3/(1/m_l+2/m_t)$ is the effective mass of the electron that ensures the sum rule $\sum_nf_{ni}=1$,\cite{Stoneham} $g_{\rm hh}=2$ is the degeneracy of the heavy-hole band and $g_s=2$ is the spin degeneracy.
%\be
%E_{\rm av}=\left<\hbar\omega\right>=\frac{\int_0^{k_0}dk_uM_r(k_u)\left[E_{\rm c}(k_u)-E_{\rm hh}(k_u)\right]}{\int_0^{k_0}dk_u\left[E_{\rm %c}(k_u)-E_{\rm hh}(k_u)\right]}
%\ee 
%is the average energy difference between the conduction and the valence band, weighted by the recombination amplitude $M_r(k_u)$.
We calculate first numerically the Coulomb scattering amplitude $M_C(k_u)$, which is shown in Fig.~\ref{MC}.
Then we calculate numerically the recombination amplitude $M_r(k_u)$, which is shown in Fig.~\ref{Mr}.
In Fig.~\ref{M} we plot the Umklapp-assisted recombination amplitude $M(k_u)$.
Inserting $M_C(k_u)$ and $M_r(k_u)$ into Eq.~(\ref{matrixelement}) and integrating over $k_u$ yields
\be
f_{\rm tot}=10.1\times 10^{-6},
\ee
which is in excellent agreement with
the oscillator strength $f_{\rm exp}=7.1\times 10^{-6}$ of the recombination of an exciton bound to a phosphorous donor in silicon reported in Ref.~\onlinecite{Dean}.
For comparison, the direct recombination of the bound electron with the bound hole over the indirect bandgap has an oscillator strength of $f_{\rm direct}=4\times 10^{-33}$. 

\section{Optical readout}
\label{readout}

In this section we show two methods that can be used to improve the optical resolution of the optical detection of the spin state of a single nucleus in silicon (see Ref.~\onlinecite{Fu}):
\begin{itemize}
\item The first method is inspired by the optically detected magnetic resonance technique (ODMR)\cite{ODMR}, where the induced ESR Rabi oscillation alters the lifetime of the bound exciton and is thus optically detectable. Since the transition from the bound hole state $M_J=-3/2$ to $M_S=+1/2$ is forbidden, mixing the spin states in an equal superposition of $M_S=+1/2$ and $M_S=-1/2$ leads to a doubling of the lifetime of the bound exciton.
\item The second method makes use of a strong ESR field that renormalizes the spin levels in such a way that the hyperfine and Zeeman splitting of the nuclear spin is increased. As in the first method, the photons of the ESR field dress the spin states of the donor electron.
\end{itemize}
The Hamiltonian for the electron-nucleus system is
\be
\cH_s=g_e\mu_B\bH\cdot\bS-g_n\mu_n\bH\cdot\bI+A\bS\cdot\bI
\ee
in the generalized rotating frame. If we choose the basis $\{\left|\uparrow_e\uparrow_n\right>,\left|\uparrow_e\downarrow_n\right>,\left|\downarrow_e\uparrow_n\right>,\left|\downarrow_e\downarrow_n\right>\}$, the Hamiltonian reads
\begin{widetext}
\be
\cH_{\rm s}=\baaaa 
\frac{g_e\mu_B}{2}H_z-\frac{g_n\mu_n}{2}H_z+\frac{A}{4} & -\frac{g_n\mu_n}{2}H_x & \frac{g_e\mu_B}{2}H_x & 0 \\
-\frac{g_n\mu_n}{2}H_x & \frac{g_e\mu_B}{2}H_z+\frac{g_n\mu_n}{2}H_z-\frac{A}{4} & -\frac{A}{2} & \frac{g_e\mu_B}{2}H_x \\
\frac{g_e\mu_B}{2}H_x & -\frac{A}{2} & -\frac{g_e\mu_B}{2}H_z-\frac{g_n\mu_n}{2}H_z-\frac{A}{4} & -\frac{g_n\mu_n}{2}H_x \\
0 & \frac{g_e\mu_B}{2}H_x & -\frac{g_n\mu_n}{2}H_x & -\frac{g_e\mu_B}{2}H_z+\frac{g_n\mu_n}{2}H_z+\frac{A}{4}
\ea.
\ee
\end{widetext}
The energy levels of the electron and nuclear spin are shown in Fig.~\ref{Boundexcitonlevels}.

\begin{figure}[htb]
\includegraphics[width=8.5cm]{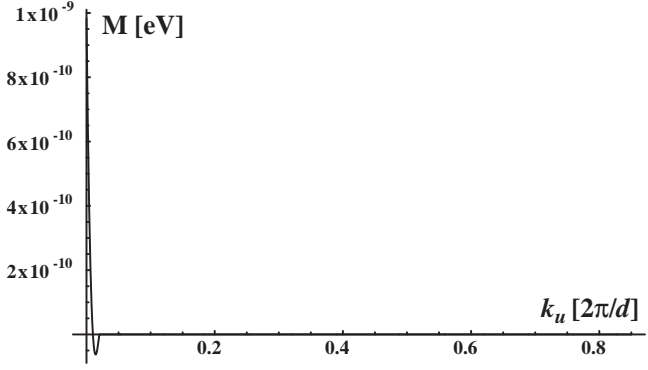}
\caption[]{Due to the large overlap of the unbound scattered electron and the hole around $k_u=0$ (see Fig.~\ref{Mr}), the Umklapp-assisted recombination amplitude $M_{\rm tot}$ is dominated by the terms $M(k_u)$ around $k_u=0$. For example, $f(k_u=0)=1.1\times 10^{-8}$.}
\label{M}
\end{figure}

Typical external magnetic fields used in experiments are $H_z=3.0,5.0$, and $7.0$ T.
In a field of $H_z=3.0$ T the Zeeman splittings of the electron and nuclear spin are $g_e\mu_BH_z=0.35$~meV and $g_n\mu_nH_z=0.21$~$\mu$eV, respectively. The hyperfine splitting is $A=0.50$~$\mu$eV. Thus the splitting between $\left|\uparrow_e\uparrow_n\right>$ and $\left|\uparrow_e\downarrow_n\right>$ is $\Delta_{\uparrow_e}=E_{\uparrow_e\uparrow_n}-E_{\uparrow_e\downarrow_n}=A/2-g_n\mu_nH_z=0.03$~$\mu$eV, whereas the splitting between $\left|\downarrow_e\uparrow_n\right>$ and $\left|\downarrow_e\downarrow_n\right>$ is $\Delta_{\downarrow_e}=E_{\downarrow_e\downarrow_n}-E_{\downarrow_e\uparrow_n}=A/2+g_n\mu_nH_z=0.46$~$\mu$eV.
The hyperfine and Zeeman splittings for $H_z=3.0,5.0$, and $7.0$ T are shown in Tab.~\ref{splittings}.

We follow here the derivation of dressed states given in Ref.~\onlinecite{Cohen-Tannoudji1992}.
The Hamiltonian of the quantized photon field is
\be
\cH_{\rm p}=\hbar\omega \hat{a}\D \hat{a},
\ee
where $\hat{a}$ ($\hat{a}\D$) is the annihilation (creation) operator of a photon.
The interaction between the microwave photons and spin of the electron of the Phosphorus donor is described by
\be
V=-\bmu\cdot\bB,
\ee
where the quantized magnetic field is given by
\be
\bB=\sqrt{\frac{\hbar}{2\epsilon_0c_p^2L^3}}\left[\hat{a}\frac{i\left(\bk\times\bepsilon\right)}{k}
+\hat{a}\D\frac{(-i)\left(\bk\times\bepsilon^*\right)}{k}\right],
\ee
where $c_p$ is the light velocity.
The spin-photon interaction can be simplified to
\be
V=g_{\rm sp}\left[\left({\bf e}\cdot\bS\right)\hat{a}+\left({\bf e}^*\cdot\bS\right)\hat{a}\D\right].
\ee
We use the circular polarization vectors
\be
{\bf e}_\pm=\frac{1}{\sqrt 2}\left({\bf e}_x\pm i{\bf e}_y\right)
\ee
and the spin ladder operators $S_\pm=S_x\pm iS_y$. Then we obtain
\bea
V_{\sigma^+} & = & \frac{g_{\rm sp}}{\sqrt 2}\left(\hat{a}S_++\hat{a}\D S_-\right), \nn\\
V_{\sigma^-} & = & \frac{g_{\rm sp}}{\sqrt 2}\left(\hat{a}S_-+\hat{a}\D S_+\right).
\eea
There are two coupled states
\bea
\left|\phi_{\downarrow_e}\right> & = & \left|\downarrow_e,N+1\right>, \nn\\
\left|\phi_{\uparrow_e}\right> & = & \left|\uparrow_e,N\right>.
\eea
If the interaction vanishes, the energies are $E_{\downarrow_e}=(N+1)\hbar\omega-\frac{1}{2}\hbar\omega_0$ and 
$E_{\uparrow_e}=N\hbar\omega+\frac{1}{2}\hbar\omega_0$.
The energy separation is $\hbar\omega_{\downarrow_e\uparrow_e}=\hbar(\omega-\omega_0)$.
The matrix elements of $V_{\sigma^+}$ read
\bea
\left<\phi_{\downarrow_e}\right|V_{\sigma^+}\left|\phi_{\downarrow_e}\right> & = & 
\left<\phi_{\uparrow_e}\right|V_{\sigma^+}\left|\phi_{\uparrow_e}\right>=0, \nn\\
\left<\phi_{\uparrow_e}\right|V_{\sigma^+}\left|\phi_{\downarrow_e}\right> & = & \frac{g_{\rm sp}}{\sqrt 2}\sqrt{N+1}
\approx \hbar\Omega=g_{\rm sp}\sqrt{\frac{\left<N\right>}{2}}.
\eea
The eigenstates are
\bea
\left|\chi_1(N)\right> & = & \sin\theta\left|\phi_{\downarrow_e}\right>+\cos\theta\left|\phi_{\uparrow_e}\right>, \nn\\
\left|\chi_2(N)\right> & = & \cos\theta\left|\phi_{\downarrow_e}\right>-\sin\theta\left|\phi_{\uparrow_e}\right>,
\eea
where $\tan2\theta=-\Omega/\left(\omega-\omega_0\right)$, $0\le 2\theta<\pi$.
The eigenenergies are
\be
E_{1/2}=\left(N+\frac{1}{2}\right)\hbar\omega\pm\hbar\sqrt{\left(\frac{\omega-\omega_0}{2}\right)^2+\left(\frac{\Omega}{2}\right)^2}.
\ee
Let us tune the oscillating transverse microwave field to the transition between $\left|\uparrow_e\uparrow_n\right>$ and $\left|\downarrow_e\uparrow_n\right>$. Then the eigenstates are
\bea
\left|\chi_{\uparrow_n1}\right> & = & \left(\left|\phi_{\downarrow_e}\right>+\left|\phi_{\uparrow_e}\right>\right)\left|\uparrow_n\right>/\sqrt 2, \nn\\
\left|\chi_{\downarrow_n1}\right> & = & \left(\sin\theta\left|\phi_{\downarrow_e}\right>+\cos\theta\left|\phi_{\uparrow_e}\right>\right)\left|\downarrow_n\right>/\sqrt 2, \nn\\
\left|\chi_{\downarrow_n2}\right> & = & \left(\cos\theta\left|\phi_{\downarrow_e}\right>-\sin\theta\left|\phi_{\uparrow_e}\right>\right)\left|\downarrow_n\right>/\sqrt 2, \nn\\
\left|\chi_{\uparrow_n2}\right> & = & 
\left(\left|\phi_{\downarrow_e}\right>-\left|\phi_{\uparrow_e}\right>\right)\left|\uparrow_n\right>/\sqrt 2,
\eea
where $\tan2\theta=-\frac{1}{2}g_e\mu_BH_x/A$.
The eigenenergies of the electron-nucleus system are
\bea
E_{\uparrow_n1} & = & -\frac{g_e\mu_B}{2}H_z-\frac{g_n\mu_n}{2}H_z-\frac{A}{4}-\frac{g_e\mu_BH_x}{2}, \nn\\
E_{\downarrow_n1} & = & -\frac{g_e\mu_B}{2}H_z+\frac{g_n\mu_n}{2}H_z
+\sqrt{\left(\frac{A}{4}\right)^2+\left(\frac{g_e\mu_BH_x}{2}\right)^2}, \nn\\
E_{\downarrow_n2} & = & \frac{g_e\mu_B}{2}H_z+\frac{g_n\mu_n}{2}H_z
-\sqrt{\left(\frac{A}{4}\right)^2+\left(\frac{g_e\mu_BH_x}{2}\right)^2}, \nn\\
E_{\uparrow_n2} & = & \frac{g_e\mu_B}{2}H_z-\frac{g_n\mu_n}{2}H_z+\frac{A}{4}+\frac{g_e\mu_BH_x}{2}.
\label{eigenvalues}
\eea
The oscillator strength for the exciton recombination in Si:P is $f=7.1\times 10^{-6}$ (see Ref.~\onlinecite{Dean}). The binding energy of the exciton to the phosophorus donor is $E_{\rm binding}=4.7$ meV.
%Taking a mode volume of $1.0$~$\mu$m$^3$, the resulting interaction energy between exciton and photon is $E_{\rm int}=0.02$~$\mu$eV. 
The recombination rate is about $w=400$~s$^{-1}$. This leads to an interaction energy of
\be
E_{\rm int}=\sqrt{\hbar w\Gamma}=12 \,{\rm neV},
\label{ep}
\ee
where $\Gamma=100$~$\mu$eV is the linewidth of the recombination.

\begin{table}[htb]
\begin{tabular}{|c|c|c|c|c|}
\hline
$H_z$ & $g_e\mu_BH_z$ & $g_n\mu_nH_z$ & $\Delta_{\uparrow_e}$ & $\Delta_{\downarrow_e}$  \\\hline
3.0 T & 0.35 meV & 0.21 $\mu$eV & 0.03 $\mu$eV & 0.46 $\mu$eV  \\
5.0 T & 0.58 meV & 0.36 $\mu$eV & -0.11 $\mu$eV & 0.61 $\mu$eV  \\
7.0 T & 0.81 meV & 0.50 $\mu$eV & -0.25 $\mu$eV & 0.75 $\mu$eV  \\\hline
\end{tabular}
\caption{The hyperfine and Zeeman splitting of the spin states of the donor electron.}
\label{splittings}
\end{table} 

If the electron-photon interaction energy $E_{\rm int}$ is much weaker than the microwave coupling energy $g_e\mu_BH_x$, the states $\left|\chi_{\uparrow_n1}\right>$, $\left|\chi_{\downarrow_n1}\right>$, $\left|\chi_{\downarrow_n2}\right>$, and $\left|\chi_{\uparrow_n2}\right>$ are good eigenstates. Then the ODMR method works, because the final state of the radiative recombination is given by $\left|\chi_{\uparrow_n1}\right>$, which doubles the lifetime of the bound exciton.
The transverse microwave field should have a strength of at least $H_x=1.0$ G, leading to a microwave coupling energy of $g_e\mu_BH_x=10.0$ neV.

\begin{figure}[htb]
\includegraphics[width=8.5cm]{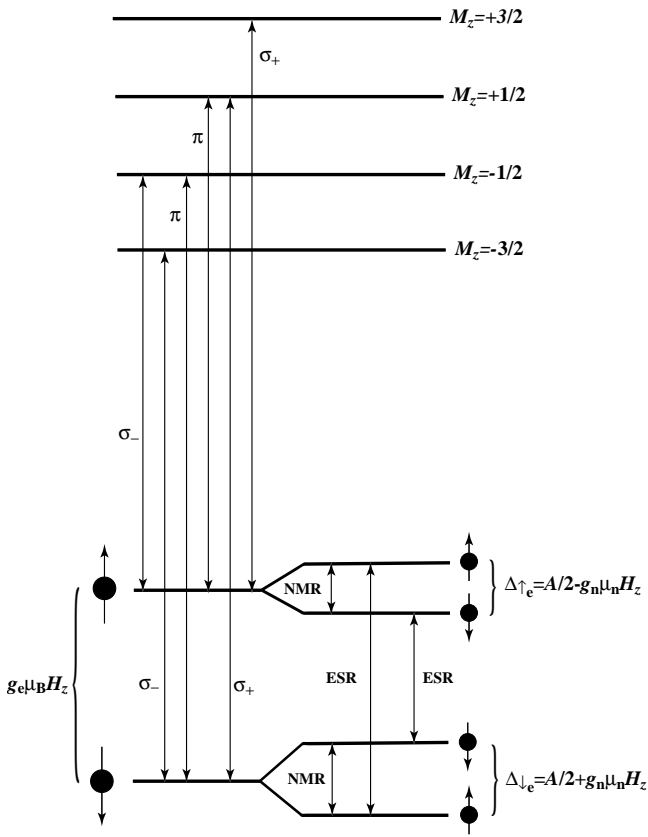}
\caption[]{The energy levels of the electron and nuclear spin of the donor electron are shown at the botton. The energy levels of the bound hole are shown at the top.}
\label{Boundexcitonlevels}
\end{figure}

In addition, it can be seen from Eq.~(\ref{eigenvalues}) that the hyperfine splittings are increased, which effectively increases the optical resolution.
In order to obtain at least a 10\% increase of the hyperfine+Zeeman splitting of the two lowest energy levels, the oscillating transverse magnetic field must have a strength of $10.0$ G, which yields the transverse Zeeman splittings of the electron and nuclear spin of $g_e\mu_BH_x=0.12$~$\mu$eV and $g_n\mu_nH_x=0.07$~neV, respectively. 
For a longitudinal magnetic field of $H_z=3.0$ T, the shifted hyperfine splitting is $\Delta_{2}=E_{\uparrow_n2}-E_{\downarrow_n2}=0.10$~$\mu$eV, whereas $\Delta_{1}=E_{\downarrow_n1}-E_{\uparrow_n1}=0.53$~$\mu$eV.
The shifted hyperfine and Zeeman splittings for $H_z=3.0,5.0$, and $7.0$ T are shown in Tab.~\ref{ODMR}.

The frequency-resolved photon detection has the advantage that a single photon detection is sufficient for determining the spin of the nucleus, whereas the ODMR measurement needs to be done in an ensemble in order to determine the radiative recombination time.
Maybe the hyperfine+Zeeman splitting of $\Delta_{\downarrow_e}=A/2+g_n\mu_nH_z=0.75$~$\mu$eV in a $H_z=7.0$ T field is already sufficient for frequency-resolved photon detection.

\begin{table}[htb]
\begin{tabular}{|c|c|c|c|c|}
\hline
$H_z$ & $\Delta_{\uparrow_e}$ & $\Delta_{\downarrow_e}$ & $\Delta_2$ & $\Delta_1$ \\\hline
3.0 T & 0.03 $\mu$eV & 0.46 $\mu$eV & 0.10 $\mu$eV & 0.53 $\mu$eV \\
5.0 T & -0.11 $\mu$eV & 0.61 $\mu$eV & -0.04 $\mu$eV & 0.68 $\mu$eV \\
7.0 T & -0.25 $\mu$eV & 0.75 $\mu$eV & -0.18 $\mu$eV & 0.82 $\mu$eV \\\hline
\end{tabular}
\caption{Change of the hyperfine and Zeeman splitting of the spin states of the donor electron.}
\label{ODMR}
\end{table}

\section{Conclusion}

We calculated the oscillator strength of recombination of an exciton bound to a donor electron in silicon. We showed that the Umklapp-assisted recombination, consisting of a scattering between the bound electron and the donor electron and the recombination of the unbound scattered electron, gives the main contribution. The calculation of the Umklapp-assisted recombination was done in second-order perturbation theory. We made use of the empirical pseudopotential method to find the Bloch wavefunctions of silicon. The calculated oscillator strength is in excellent agreement with the experiment. We gave also two methods to improve the resolution of the detection of a nuclear spin of a donor electron in silicon.

\begin{acknowledgements} 
We thank Marilyn Hawley, Geoff Brown, and Holger Grube for useful discussions. This work has been supported by LDRD from Los Alamos National Laboratory and the US NSF DMR-0403465.
\end{acknowledgements}

\appendix

\section{Matrix element of Umklapp-assisted recombination}
\label{matrix}

We are going to calculate the matrix elements $M_{\rm C}$ $M_{\rm r}$ in the reciprocal lattice space by Fouriertransforming the wavefunctions.
The Bloch wavefunctions can be expanded in reciprocal lattice vectors as
\be
\phi_{{\rm B}\bk_{z0}}(\br)=e^{ik_{z0}z}\sum_{\bG}u_{c\bG}e^{i\bG\cdot\br}
\ee
for the electron wavefunctions and
\be
\phi_{{\rm h}}(\br)=\sum_{\bG}u_{{\rm h}\bG}e^{i\bG\cdot\br}
\ee
for the hole wavefunction. Thus the electron wavefunctions in $\bk$-space are given by
$\tilde{\psi}_{\rm e}(\bk,k_c)=\sum_{\bG}u_{c\bG}\tilde{\psi}_{{\rm e}\bG}(\bk,k_c)$ with
\bea
\tilde{\psi}_{{\rm e}\bG}(\bk,k_c) & = & \left(\frac{2}{\pi a^2b}\right)^{\frac{3}{2}}
\frac{1}{\left(\frac{1}{a^2}+(k_x+G_x)^2\right)} \nn\\
& & \times\frac{1}{\left(\frac{1}{a^2}+(k_y+G_y)^2\right)} \nn\\
& & \times\frac{1}{\left(\frac{1}{b^2}+(k_z+G_z-k_c)^2\right)},
\eea
centered at $k_c$,
and the hole wavefunction by $\tilde{\psi}_{\rm h}(\bk)=\sum_{\bG}u_{h\bG}\tilde{\psi}_{{\rm h}\bG}(\bk)$ with
\bea
\tilde{\psi}_{{\rm h}\bG}(\bk) & = & \left(\frac{2}{\pi c^3}\right)^{\frac{3}{2}}
\frac{1}{\left(\frac{1}{c^2}+(k_x+G_x)^2\right)} \nn\\
& & \times\frac{1}{\left(\frac{1}{c^2}+(k_y+G_y)^2\right)} \nn\\
& & \times\frac{1}{\left(\frac{1}{c^2}+(k_z+G_z)^2\right)}.
\eea
So the initial bound and donor electron wavefunctions are $\tilde{\psi}_{\rm B}(\bk)=\tilde{\psi}_{\rm D}(\bk)=\tilde{\psi}_{\rm e}(\bk,k_0)$.
The intermediate scattered electron wavefunction in $\bk$-space is given by
$\tilde{\psi}_{\rm B}'(\bk')=\sum_{\bG}u_{c\bG}\tilde{\psi}_{{\rm B}\bG}(\bk')$ with
\bea
\tilde{\psi}_{{\rm B}\bG}(\bk') & = & \left(\frac{2}{\pi a^2b}\right)^{\frac{3}{2}}
\frac{1}{\left(\frac{1}{a_u^2}+(k_x+G_x)^2\right)} \nn\\
& & \times\frac{1}{\left(\frac{1}{a_u^2}+(k_y+G_y)^2\right)} \nn\\
& & \times\frac{1}{\left(\frac{1}{a_u^2}+(k_z+G_z-k_u)^2\right)},
\eea
where $a_u=1000d$,
the intermediate donor electron wavefunction by
$\tilde{\psi}_{\rm D}'(\bk')=\tilde{\psi}_{\rm e}(\bk,k_{00})$,
where $k_{00}=1.15k_{\rm si}$,
and the intermediate hole wavefunction by $\tilde{\psi}_{\rm h}'(\bk')=\tilde{\psi}_{\rm h}(\bk)$.
Note that only the bound and the donor electron scatter off each other.

Since the Coulomb interaction is local within each Brillouin zone, interference effects can be neglected and thus it is sufficient to calculate the Coulomb scattering matrix element within the first Brillouin zone, i.e.
\bea
M_{\rm C} & = & \left<\psi_{\rm B}'\psi_{\rm D}'\left|V_{\rm C}\right|\psi_{\rm B}\psi_{\rm D}\right>
= \frac{q^2}{\epsilon}\int d^3k_{\rm B}'\int d^3k_{\rm B}\int d^3k_{\rm D} \nn\\
&  & \times 
\tilde{\psi}{'}_{{\rm B}0}^*(\bk_{\rm B}')\tilde{\psi}{'}_{{\rm D}0}^*(\bk_{\rm B}-\bk_{\rm B}'+\bk_{\rm D})
\frac{1}{\left(\bk_{\rm B}-\bk_{\rm B}'\right)^2+\xi^2} \nn\\
& & \times\tilde{\psi}_{{\rm B}0}(\bk_{\rm B})\tilde{\psi}_{{\rm D}0}(\bk_{\rm D}),
\eea
where we used the relation
\bea
\lefteqn{
\int d^3r_{\rm B}\int d^3r_{\rm D} e^{-i\bk_{\rm B}'\cdot\br_{\rm B}}e^{-i\bk_{\rm D}'\cdot\br_{\rm D}}
\frac{q^2e^{-\xi r_{\rm BD}}}{4\pi\epsilon r_{\rm BD}}
e^{i\bk_{\rm B}\cdot\br_{\rm B}}e^{-i\bk_{\rm D}\cdot\br_{\rm D}}
} \nn\\
& = & \frac{q^2}{(2\pi)^3\epsilon}\int d^3k_{\rm BD}
\frac{\delta(\bk_{\rm B}-\bk_{\rm B}'+\bk_{\rm BD})\delta(\bk_{\rm D}-\bk_{\rm D}'-\bk_{\rm BD})}{k_{\rm BD}^2+\xi^2}
\nn\\
& = & \frac{q^2}{(2\pi)^3\epsilon}\frac{1}{\left(\bk_{\rm B}-\bk_{\rm B}\right)^2+\xi^2}
\delta(\bk_{\rm B}-\bk_{\rm B}'+\bk_{\rm D}-\bk_{\rm D}'),
\eea
where $\br_{\rm BD}=\br_{\rm B}-\br_{\rm D}$ and $\bk_{\rm BD}=2\pi/\br_{\rm BD}$.
Numerical calculations show that the Coulomb potential is well approximated by 
\be
\tilde{V}_C(\bk_{\rm B}-\bk_{\rm B}')\approx\frac{q^2}{\epsilon}
\frac{1}{\left(k_{{\rm B}z}-k'_{{\rm B}z}\right)^2+\xi^2},
\ee
i.e. only the longitudinal dependence of the Coulomb potential in $z$ direction is taken into account. Then the integrations in $x$ and $y$ directions can be solved analytically, which yields
\bea
M_{\rm C} & = & \tau_\bot^2 \frac{q^2}{\epsilon}\int dk'_{{\rm B}z}\int dk_{{\rm B}z}\int dk_{{\rm D}z}
\tilde{\psi}{'}_{{\rm B}0}^*(k'_{{\rm B}z}) \nn\\
&  & \times 
\tilde{\psi}{'}_{{\rm D}0}^*(k_{{\rm B}z}-k'_{{\rm B}z}+k_{{\rm D}z})
\frac{1}{\left(k_{{\rm B}z}-k'_{{\rm B}z}\right)^2+\xi^2} \nn\\
& & \times\tilde{\psi}_{{\rm B}0}(k_{{\rm B}z})\tilde{\psi}_{{\rm D}0}(k_{{\rm D}z}),
\eea
where
\bea
\tau_\bot & = & \frac{1}{4\pi^2a}\left\{i\left[\log(-ia)-\log(ia)\right]\left[\log(-i\frac{a}{3})-\log(i\frac{a}{3})\right]\right. \nn\\
& & \times\left[\log(-i\frac{a}{2})-\log(i\frac{a}{2})+\log(-ia)-\log(ia)\right] \nn\\
& & +i\left[\log(-ia)-\log(ia)\right]^2\left[\log(-i\frac{a}{2})-\log(i\frac{a}{2})\right. \nn\\
& & \left.\left.-7\log(-ia)+7\log(ia)\right]\right\}.
\eea
Since the scattered electron state $\tilde{\psi}{'}_{{\rm B}0}^*(k'_{{\rm B}z})$ is unbound, we can treat it as the square root of a delta-function with width $a_u$, i.e. $\tilde{\psi}{'}_{{\rm B}0}^*(k'_{{\rm B}z})=\sqrt{\delta^{(a_u)}(k'_{{\rm B}z}-k_u)}$.
This means that we can approximate the integral $\int dk'_{{\rm B}z} \tilde{\psi}{'}_{{\rm B}0}^*(k'_{{\rm B}z})h(k'_{{\rm B}z})$ by $h(k_u)\zeta$, where 
$\zeta=\int dk'_{{\rm B}z} \tilde{\psi}{'}_{{\rm B}0}^*(k'_{{\rm B}z})$, for any function $h(k'_{{\rm B}z})$.
Thus we obtain
\bea
M_{\rm C} & = & \tau_\bot^2\zeta \frac{q^2}{\epsilon}\int dk_{{\rm B}z}\int dk_{{\rm D}z}
\tilde{\psi}{'}_{{\rm D}0}^*(k_{{\rm B}z}-k_u+k_{{\rm D}z}) \nn\\
&  & \times 
\frac{1}{\left(k_{{\rm B}z}-k_u\right)^2+\xi^2} \tilde{\psi}_{{\rm B}0}(k_{{\rm B}z})\tilde{\psi}_{{\rm D}0}(k_{{\rm D}z}),
\eea
which is solved numerically (see Fig.~\ref{MC}).

In the case of the recombination we cannot neglect interference effects, because the electron-photon interaction is nonlocal in $\bk$-space. We obtain
\be
M_{\rm r}=\frac{\hbar qA_\bot}{m}\int d^3k_{\rm h}'\tilde{\psi}_{\rm h}{'}^*(\bk_{\rm h}')k_{{\rm h}\bot}'
\tilde{\psi}_{\rm B}'(\bk_{\rm h}'),
\ee
which is solved numerically (see Fig.~\ref{Mr}).

\section{Empirical Pseudopotential Method}
\label{pseudo}

We give here a brief review of the empirical pseudopotential method.
At each lattice site there is an atom with a nucleus, core electrons, and valence electrons.
The attractive nuclear potential $V_{\rm n}$ is large and varies strongly throughout the lattice.
The main observation is that $V_{\rm n}$ is almost canceled by the repulsive potential $V_{\rm rep}$ produced by the core electrons.
So we need to consider only the valence electrons moving in a net weak one-electron potential $V_{\rm p}$, which is called the pseudopotential.

The full Bloch wavefunctions can be expressed as
\be
\Phi=\phi+\sum_t\alpha_t\varphi_t.
\ee
$\Phi$ must be orthogonal to the core wave functions $\varphi_t$, i.e. $\left<\varphi_t|\Phi\right>=0$, which yields
\be
\alpha_t=-\left<\varphi_t|\phi\right>.
\ee
Then applying the Hamiltonian $\cH_{\rm n}=p^2/2m_e+V_{\rm n}$ to $\Phi$ leads to the Schr\"odinger equation
\be
\left(\frac{p^2}{2m_e}+V_{\rm n}+V_{\rm rep}\right)\phi=E\phi,
\label{Schroedinger}
\ee
where the short-range non-Hermitian repulsive potential is given by
\be
V_{\rm rep}\phi=\sum_t\left(E-E_t\right)\varphi_t\left<\varphi_t|\phi\right>,
\ee
with $E$ being the full energy eigenvalue of $\Phi$. $\phi$ is called the pseudowavefunction.
Since the crystal potential is periodic, the pseudopotential $V_{\rm p}=V_{\rm n}+V_{\rm rep}$
and can be expanded in a Fourier series over the reciprocal lattice vectors $\bG$, i.e.
\be
V_{\rm p}=\sum_\bG \tilde{V}_0(\bG)e^{i\bG\cdot\br},
\ee
where
\be
\tilde{V}_0(\bG)=\frac{1}{L^3}\int d^3r V_0(\br)e^{-i\bG\cdot\br}.
\ee
For zincblende and diamond lattices usually a two-atom basis is chosen, such that
\be
V_0(\br)=V_{\rm cation}(\br-\tau)+V_{\rm anion}(\br+\tau),
\ee
where $\btau=(1,1,1)d/8$. Then the Fourier potential reads
\be
\tilde{V}_0(\bG)=e^{i\bG\cdot\btau}\tilde{V}_{\rm cation}(\bG)+e^{-i\bG\cdot\btau}\tilde{V}_{\rm anion}(\bG).
\ee
The Fourier coefficients can be rewritten in terms of the symmetric and antisymmetric form factors
$\tilde{V}_S=\tilde{V}_{\rm cation}+\tilde{V}_{\rm anion}$ and $\tilde{V}_A=\tilde{V}_{\rm cation}-\tilde{V}_{\rm anion}$. Thus
\be
\tilde{V}_0(\bG)=\cos(\bG\cdot\btau)\tilde{V}_S(\bG)+i\sin(\bG\cdot\btau)\tilde{V}_A(\bG),
\ee
where the prefactors of the form factors are the structure factors.
Since silicon has a diamond lattice, $\tilde{V}_A(\bG)=0$.

The pseudowavefunction can also be expanded in a Fourier series, i.e.
\be
\phi(\br)=e^{i\bk\cdot\br}u(\br)=e^{i\bk\cdot\br}\sum_\bG u_\bG e^{i\bG\cdot\br}.
\ee
Inserting the pseudowavefunction $\phi$ and the pseudopotential $V_{\rm p}$ into Eq.~(\ref{Schroedinger}) yields
\be
\frac{\hbar^2(\bk+\bG)^2}{2m_e}u_\bG+\sum_{\bG'}V_0(|\bG-\bG'|)u_{\bG'}=Eu_\bG.
\ee
Diagonalization of this Hamiltonian yields an effective mass Hamiltonian, which is the first term on the right side of Eq.~(\ref{Hamiltonian}).

\end{document}